\documentclass[structabstract,rnote]{aa}  
\usepackage{graphicx}
\usepackage{txfonts}

\begin{document}

   \title{HAT-P-13: a multi-site campaign to detect the transit of 
   the second planet in the system}

\titlerunning{New transit observations of HAT-P-13bc}
\authorrunning{Szab\'o et al}

   \author{Gy.M. Szab\'o \inst{1,2,3} \and L.L. Kiss\inst{1,4} \and J. M. 
   Benk\H{o}\inst{1} \and Gy. Mez\H{o}\inst{1}
   \and J. Nuspl\inst{1} \and Zs. Reg\'aly\inst{1} \and K. 
   S\'arneczky\inst{1} \and A. E. Simon\inst{1,2} \and G. 
   Leto\inst{5} \and R. Zanmar 
   Sanchez\inst{5} \and C.-C. Ngeow\inst{6} \and Zs. K\H{o}v\'ari\inst{1} \and R. Szab\'o\inst{1}
          }

   \institute{Konkoly Observatory of the Hungarian Academy of Sciences, PO Box 67, H-1525 Budapest, Hungary\\
              \email{szgy@konkoly.hu, kiss@konkoly.hu}
         \and
             Department of Experimental Physics and Astronomical Observatory, University of Szeged, H-6720 Szeged, Hungary
             \and
Hungarian E\"otv\"os Fellow, Department of Astronomy, University of Texas at Austin,
1 University Station, Austin, TX 78712, USA. 
             \and
             Sydney Institute for Astronomy, School of Physics A28, University of Sydney, NSW 2006, Australia
               \and
            INAF - Osservatorio Astrofisico di Catania, Via Santa Sofia 78, I-95123 Catania, Italy
            \and
            Graduate Institute of Astronomy, National Central University, No. 300, Jhongda Rd, Jhongli City, Taoyuan County 32001, Taiwan
             }

   \date{Received ...; accepted ...}

 
  \abstract
   {}
   {
   A possible transit of HAT-P-13c has been predicted to occur on 2010 
   April 28. Here we report on the results of a multi-site campaign 
   that has been organised to detect the event.}
   {CCD photometric observations have been carried out at five 
   observatories in five countries. We reached 30\%{}
   time coverage in a 5 days interval centered on the suspected transit
   of HAT-P-13c. Two transits of HAT-P-13b were also observed.}
   {No transit of HAT-P-13c has been detected while the campaign was on.
   By a numerical experiment with 10$^5$ model systems
   we conclude that HAT-P-13c is not a transiting exoplanet with a significance
   level from 65\%{} to 72\%{}, depending on the planet parameters and 
   the prior assumptions. We present two 
   times of transit of HAT-P-13b ocurring at BJD 2455141.5522$\pm$0.0010 and BJD 2455249.4508
   $\pm$ 0.0020. The TTV of HAT-P-13b is consistent with zero within 0.001 days.
   The refined orbital period of HAT-P-13b is 2.916293$\pm$0.000010 days.}
   {}


   \maketitle

\section{Introduction}

Multiple planetary systems analogous to our Solar System
play a key role in understanding planet formation and evolution. 
If planets in multiple systems display
transits as well {     (e.g. Kepler-9, Holman et al. 2010)}, 
a very detailed analysis becomes possible, resulting in
a set of dynamical parameters; and even the internal density distribution 
of the planets (Batyigin et al. 2009). 
As of this writing, three multiple systems with a transiting component
have been discovered. The CoRoT-7 system has two orbiting super-Earths, one
showing transits (L\'eger et al. 2009, Queloz et al. 2009); HAT-P-7 hosts a hot
Jupiter in a polar or retrograde orbit and a long-period companion that can
either be a planet or a star (P\'al et al. 2008, Winn et al. 2009). But the 
most 
prominent example of such systems is HAT-P-13 (Bakos et al. 2009, Winn et al. 
2010). The central star of this system
is a G4 dwarf with 1.22 M$_\odot$ mass and 1.56 R$_\odot$ radius. HAT-P-13b
is a 0.85 M$_J$ hot Jupiter on a 2.9 day orbit that has almost been
circularized. HAT-P-13c has a minimum mass of M$\sin i$=15.2 M$_J$ in a 428
day orbit with 0.69 eccentricity. Winn et al. (2010) predicted a possible transit
for the second planet, which, if confirmed, would make HAT-P-13 an extremely 
special system.

In multiple planetary systems, the most important question is whether the 
orbital planes are aligned. 
If this is the case for HAT-P-13 b and c, the exact mass of companion 
c can be derived. The $\Delta i$ mutual inclination may be derived from the 
Transit Timing
Variations of HAT-P-13b (Bakos et al. 2009). A more stringent 
constraint on coplanarity would be delivered
if HAT-P-13c also transits. In this case the coplanarity is highly probable, and
the radius and the orbit of planet c can be measured. 
If the apsides are also aligned, 
tidal dynamics can reveal planet b's internal structure, which is a
fascinating opportunity to extract unique information
on an exoplanet (Batygin et al. 2009, Fabricky 2009).

It has been unknown whether HAT-P-13c transits. Dynamical models of Mardling
(2010) suggest that the HAT-P-13 system is likely to be close to prograde 
coplanar or have a mutual inclination between 130$^\circ$ and 135$^\circ$. 
She interpreted the system geometry 
as a result of early chaotic interactions. A hypothetical 
d companion has been invoked at the early stages
of evolution {     that should have escaped later and could explain the 
vivid scattering history}.
Her argument
for coplanarity is that lower masses are favoured because of dynamical reasons,
although c's high inclination itself favours a large mutual inclination.
Winn et al. (2010) points to the observed small stellar obliquity $\psi_{*,b}$
as an indirect evidence of orbital alignment: in Mardling's model, after
having planet d escaped, 
$\psi_{*,b}$ oscillates about a mean value of $\Delta i$. Thus, 
observing small value for $\psi_{*,b}$ at any time, e.g. now, is unlikely
unless $\Delta i$ is small.

\begin{table*}
\caption{}
\centering\begin{tabular}{lllll}
\hline
Code & Telescope &  CCD & FoV & resolution \\
\hline
K60 & Konkoly 0.6 Schmidt, Piszk\'estet\H{o}, Hungary & 1526$\times$1024 KAF & 25$^\prime\times$17$^\prime$ & 1.0$^{\prime \prime}$/pixel \\

K100 & Konkoly 1.0 RCC,  Piszk\'estet\H{o}, Hungary & 1340$\times$1300
PI VersArray 1300b NTE& 7$^\prime \times$7$^\prime$ & 
0.32$^{\prime\prime}$/pixel\\

SLN & INAF-OACt 0.91, Fracastoro, Italy & 1100$\times$1100 KAF1001E & 13$^\prime \times$13$^\prime$ & 0.77 $^{\prime \prime}$/pixel\\

TEN & 0.8 RCC Tenagra II, Arizona, USA & 1024$\times$1024 & 14.8$^\prime \times$14.8$^\prime$ & 0.81 $^{\prime \prime}$/pixel \\

LNO & Langkawi 0.5 RCC, Malaysia & 1024$\times$1024 SBIG 1001E CCD & 20$^\prime \times$20$^\prime$ & 1.2 $^{\prime\prime}$/pixel\\

SLT & Lulin 0.4 RCC, Taiwan & 3056$\times$3056 Apogee U9000 & 50.7$^\prime\times$50.7$^\prime$ & 0.99$^{\prime\prime}$/pixel\\

\hline
\end{tabular}
\end{table*}

\begin{table*}
\caption{Observations during the HAT-P-13c campaign. Telescope codes: 
K60: Konkoly 60 cm Schmidt, TEN: Tenagra, SLT: Lulin, LNO: Langkawi,
SLN: INAF-OACt. 
Observation windows and the number of photometry points are indicated.}
\centering\begin{tabular}{rllllll}
\hline
Date ~~~& K60 & TEN & SLT & LNO & SLN\\
\hline
2010--04--22& 20:23--22:23 (80)& &\\
04--25& 18:40--23:33 (190)~~~& 03:01--05:17 (101)~~~ & 12:17--15:03 (108)~~~ & \\
04--26& 18:43--21:27 (134)& 04:44--07:01 (100) & & 14:28--14:56 (34)\\
04--27 && 03:10--05:26 (85)    &&& 20:46--23:56 (162)\\
04--28& 18:41--22:53 (139)& 05:30--06:52 (61)& &13:34--14:55 (23)& 20:37--00:08 (116)\\
04--29& 18:45--23:15 (349)& 04:55--06:49 (84)& & 13:42--15:32 (58)& 20:53--23:58 (128)\\
04--30& 18:43--23:17 (342)& 04:55--06:55 (51)&\\
05--01& 19:21--20:32 (66)&  05:39--06:17 (45) & 11:48--14:16 (55)&12:42--14:45 (60)\\
05--03& 19:25--22:50 (252)& &&\\
\hline
\end{tabular}
\end{table*}

\begin{figure*}
\begin{centering}
\includegraphics[width=17.5cm]{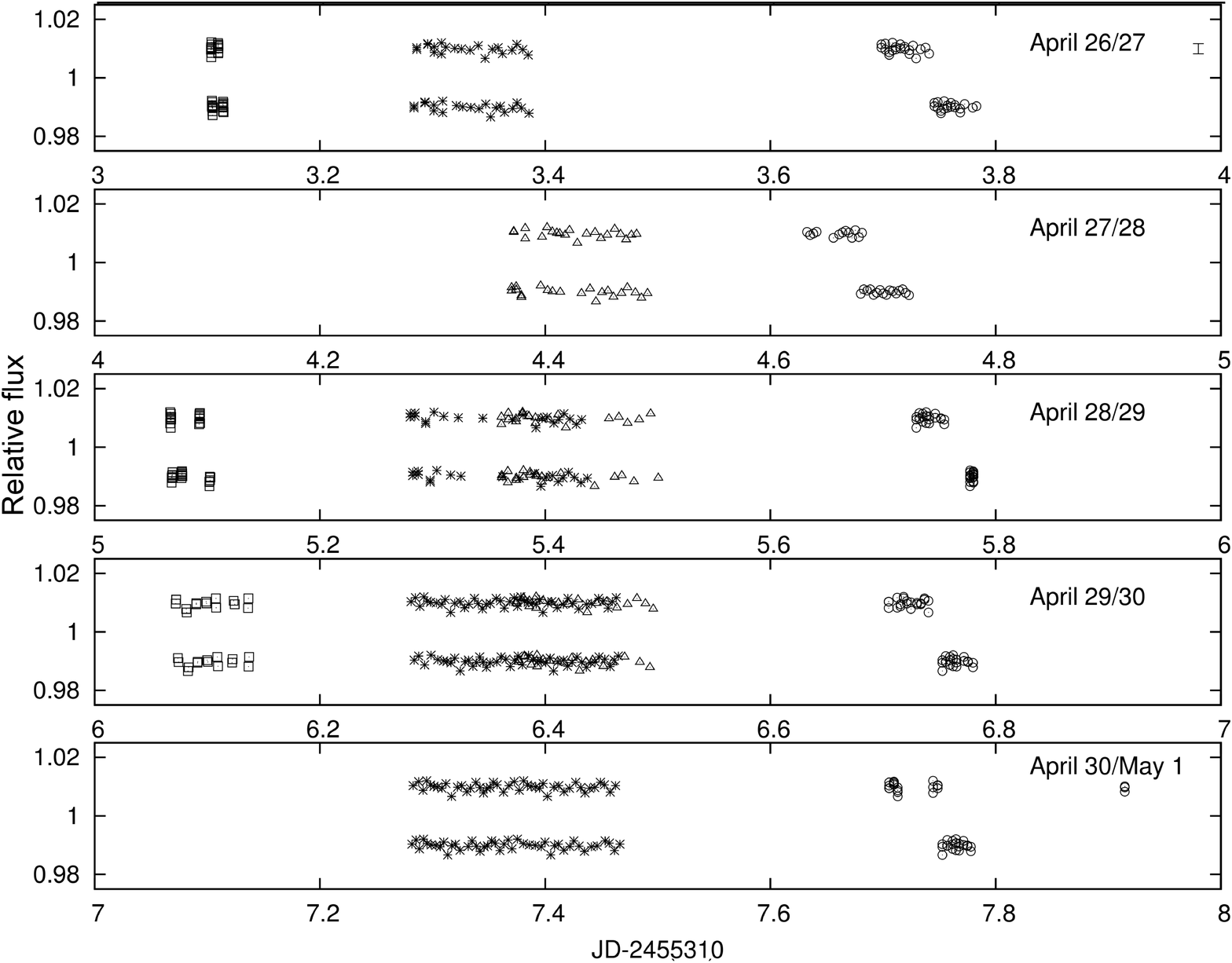}
\caption{Observations of HAT-P-13 between April 26--30. 
Observations are shifted with $+$0.01 (V points) and $-$0.01 (R points) as
indicated. Different symbols are applied for the different
observatories: square: Langkawi, stars: Konkoly, triangles: INAF-OACt, circle:
Tenagra. The typical standard deviation is 0.0013 in R and 0.0014 in V. 
A $\pm0.0015$ error bar is indicated in the upper right corner of the top panel.
}
\end{centering}
\end{figure*}

The refined orbital elements suggested
that the transit - if it happened - should have occurred around 2010 April 28, 17 UT,
(JD 2455315.2) with 1.9 days 
FWHM of transit probability and a maximal duration of 14.9 hours
(Winn, 2010). We started monitoring of HAT-P-13 for further transits
in November 2009 and organised an international campaign in the 2 weeks
surrounding the expected transit {     of HAT-P-13c}.


%

\section{Observations and Data Reduction}

The seasonal visibility of HAT-P-13 is quite unfavourable in April.
Hence the longest possible run at mid-northern latitudes 
may last 3--4 hours after twilight with 
observations ending at high (X$>$2)
airmass. Our data were collected at 5 observing sites with 6 telescopes, 
and, due to the weather conditions, 
30\%{} time coverage was reached. The telescope parameters and the log of the 
observations is shown in
Table 1 and 2, respectively.

The observing strategy was the same in most observatories: a sequence of
RRRVVV was repeated continuously, while Tenagra Observatory measured the first
half of the light curve in R, and the second half in V.
Integration time was adjusted all 
along the night to compensate for the air mass variation in 
an effort to take advantage of the full dynamic range of the 
camera. The average exposure time was about 65 s and 35 s 
in the V and R bands, respectively. Each night several bias, dark and sky 
flat images were taken for calibration.

Before the multisite campaign, we observed HAT-P-13 on 8 additional nights.
Two nights (2009-11-05/06 and 2010-02-21/22) included a transit of
HAT-P-13b, the rest acquired as out-of-transit observations.
{    
In these observations, the K100 telescope was 
also involved. 
No transit signal exceeding a depth of 0.005 (3-sigma level) 
was observed during the following out-of-transit observation runs: 
2009--11--05/06, 23:03--03:45 UT (1 RCC),
2010--01--11/12, 01:41--04:29 (1 RCC),
2010--01--14/15, 21:41--23:19 (0.6 Schmidt),
2010--01--16/17, 22:20--03:39 (0.6 Schmidt),
2010--02--21/22,  18:32--02:19 (0.6 Schmidt),
2010--03--18/19,  19:08--00:03 (0.6 Schmidt),
2010--03--18/19,  20:08-23:37 (1 RCC)
2010--03--19/20,  21:38--00:11 (0.6 Schmidt),
2010--03--28/29,  18:30--00:16 (0.6 Schmidt).

Transits of HAT-P-13b were} analyzed with an automated image processing and aperture 
photometry pipeline developed in 
{\sc gnu-r}\footnote{r-project.org} 
environment. {     The flat image was constructed as the median of the normailzed flat frames (i.e. each acquired images were divided by the mean of their pixel values), and that similar procedures were performed for darks and bias.}
After the standard calibrations, star identification
was performed. Comparison stars were selected iteratively 
for attaining the best S/N in the light curve. Finally, 
3 comparison stars were used in all images {     (2MASS J08392449+4723225, 2 MASS J08392164+4720500, 2MASS J08391779+4722238), to ensure the consistency of the
entire dataset. $J-K$ colors of the comparison stars are 0.419, 0.384 and 0.337,
quite close to $J-K=0.353$ of HAT-P-13.}

The data were corrected for systematics with {     
the well-known parameter decorrelation technique (e.g. Robinson et al. 1995), in our
case applying the specific implementation of the 
External Parameter Decorrelation (EPD) in constant mode (Bakos et al. 2010). The observed external parameters
were the PSF of stellar profiles and the local photometry of the 
flat field image at the same $X,Y$ position where the stars were 
observed. The variation of stellar profile is a known error source 
which has been involved in most standard reduction pipelines of 
exoplanet photometry. Considering the flat field image intensities 
as an error source means assuming that dividing with the flat 
field under/overestimates the neccessary correction by a factor of a 
few 0.1\%{}. We experienced that most of the artificial patterns of the light curves is due to systematic residuals of flat field correction and could be well eliminated this way.} In the end, 6585 raw photometric points
were extracted. We omitted points out of the 5--95\%{} 
quantile interval {     of the measured fluxes} 
and averaged the surviving points by 3. 
This resulted in 1952 data points submitted to further analysis.

\section{Results}

\subsection{Significance analysis of the null detection}

In Fig. 1 we plot sample light curves from the multisite campaign. 
The panels show the combined light curves from April 26, 27, 28, 29 
and 30. Neither signs of ingress or egress nor
significant deviations from the average brightness have been observed. 
These 
features strongly suggest that all observations are out of 
transit, and HAT-P-13c is
likely to be a non-transiting exoplanet.

What is the significance of this conclusion? The time coverage
of our data is 30\%. Thus the first answer could be that a
transit could happen anytime in 70\% of the time, i.e. when observations were
not done, and this null result is essentially insignificant. 
But this conclusion is not correct and in fact, 
our observations rule out the majority of transiting orbits for HAT-P-13c.

We did a numerical experiment to determine the
quantitative measure of the significance.
A set of
$10^5$ exoplanets were simulated on a similar orbit to HAT-P-13 (428 days 
period around an 1.22~R$_\odot$, 1.56~R$_\odot$ star). 
{    
The radius of the planet was assumed to be 1.2 $R_J$, {     which is the typical size
for} the most massive known exoplanets. With this choice, the density of 
HAT-P-13c is 8.7 times of the Jupiter.  
The orbital eccentricity of the model was $e=0.691$, the argument of periastron was
$\omega=176.7^\circ$, coefficients for quadratic limb darkening were
$\gamma_1=0.3060$, $\gamma_2=0.3229$ (planet and orbit parameters 
from Bakos et al. 2009).
To include grazing transits, the value of the impact 
parameter $b$ was allowed to be $>1$ and was drawn from an uniform distibution between 0 and 1.08. The transit time followed a uniform distribution in the April 26.5 UT and April 30.5 UT interval.} 
{     In some possible planet configurations it is probable that data of a given run could have included only the bottom of the transit. This should be seen as a slight offset from the rest of the runs, but that this cannot be detected because of non-photometric conditions.} 
What we are sure about is that ingress and egress phases were not detected
within our time coverage. Solely this information constrains the possible
orbits seriously in the transit time--impact parameter space.

{     Model transit light curves were sampled at the times of observation points (all data in Table 2), sorted to observation runs and the average level was individually subtracted. We added bootstrap noise to the individual points (the measured light curve errors were randomly added to the simulated values with subtitution). Then a $\chi^2$ test was applied to check whether the simulations are inconsistent with zero at the 99\% significance level. This way we identified those configurations of HATP-P-13c which should have been observed in our measurements (we call these observable configurations in the following). Because our observations are consistent with zero variation, observable configurations are explicitelly excluded by our data.

We identified that 72\%{} of the $10^5$ model transit configurations would have been
observable. Therefore the hypothesis of HAT-P-13c to be a transiting exoplanet can be rejected with 72\%{} confidence. By allowing the mean transit times to be distributed
normally around April 28 17 UT with 1.9 days standard deviation, the level
of significance turns out to be 70\%{}. The level of significance does not vary significantly in the range of orbits allowed by the parameter uncertainties in Bakos et al (2009), because the errors are rather small (3\%{} in $e$ and 0.3\%{} in $\omega$). We reduced the model light curves in amplitude to define the size limit where the detection efficiency starts decreasing significantly. The resulting significance was $65\%{}$ when the amplitude was reduced by 0.45. The planet size corresponding to this signal amplitude is 1.04 $R_J$, which is our detection limit. The conclusion is that roughly
three quaters of all possible transiting configurations are excluded by our observations.
}

This result does not mean that HAT-P-13c could not orbit on an aligned orbit
with HAT-P-13b. HAT-P-13c is quite far from the central star, 
hence the star's apparent diameter
is 0.6 degrees as seen from the planet. 
Thus, transiting configurations require the orbit to be in a thin region,
very close to our line of sight. There is a huge set of configurations
with HAT-P-13c on {     an orbit close to that of planet b,} without displaying any transits. 
In this case, Transit Timing Variation (TTV) {     of HAT-P-13b} can reveal the orientation
of HAT-P-13c's orbital plane (Bakos et al. 2009).

\subsection{Transit Timing Variations of HAT-P-13b}

\begin{figure}
\includegraphics[bb=138 250 432 488, width=8cm]{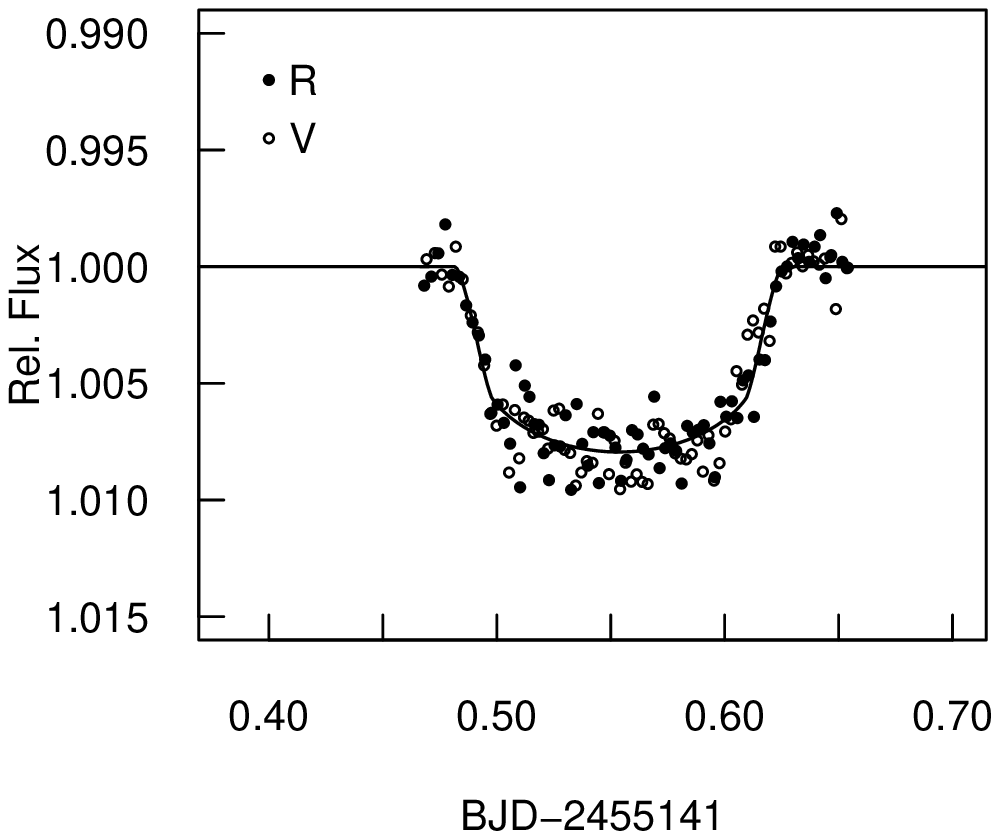}
\includegraphics[bb=138 250 432 488, width=8cm]{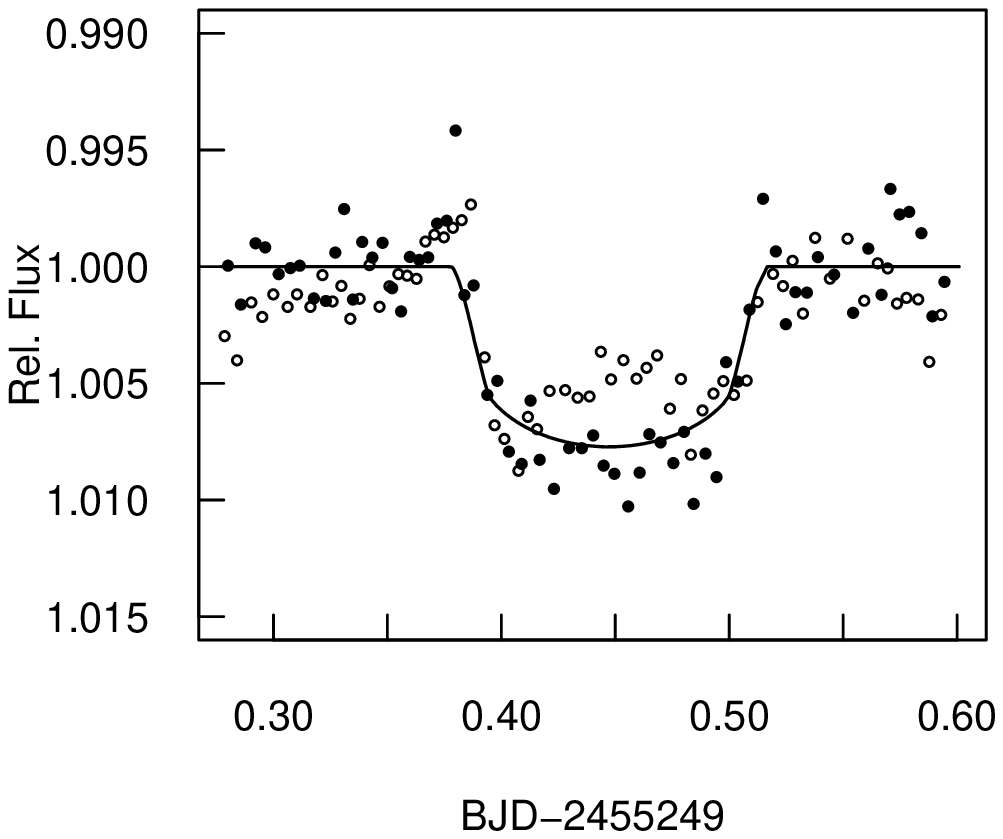}
\caption{Model fit to the transit on November 05/06, 2009 (upper panel) 
and February 21/22, 2010 (lower panel). V and R band data 
are plotted with open and soild dots, respectively.}
\end{figure}

Before the suspected transit of HAT-P-13c, two transits of HAT-P-13b were
observed to refine the period and to search for Transit
Timing Variations (TTV). Data 
from 2009-11-05/06 {     (measured with the K100 telescope, Table 1)} and 
2010-02-21/22 {     (K60 telescope)} are plotted 
in Fig 2.  In November (upper panel in Fig. 2), 
the sky was photometric
during the transit, but it was foggy in the evening and from
40 minutes after the egress phase. In February, 2010, 
cirri were present
that significantly affected the V band data, but the R light curve
was well reconstructed with constant EPD (see lower panel in Fig. 2).

Times of minima were determined by fitting a model light curve, similarly
to Szab\'o et al. (2010). 
For the November 2009 transit, both V and R data were included in the fitting,
while we used only the R curve for the February 2010 transit. (However,
even including the more noisy V curve does not change the mid-transit time
by more than 0.0004 days.)

To reduce the degrees of freedom in the fit, the shape of the model 
was not adjusted; we used previously published parameters (Winn et al. 2010). 
The model light curve was calculated with our transit simulator
(Simon et al. 2009, 2010). The model 
was shifted in time, minimising the rms scatter of the measurements. 
We determined new transit times as: BJD 2455141.5522$\pm$0.001 and
2455249.4508$\pm$0.002. Seven transit times were published by 
Bakos et al. (2009) which were included in the TTV analysis.
Combining all data, we refined the period of HAT-P-13b to be 
2.916293$\pm$0.000010 days, while the determined TTV diagram
is plotted in Fig. 3. 
All points are consistent with
zero within the error bars.
It has to be noted that HAT-P-13b must exhibit some TTV,
because of the perturbations by HAT-P-13c. 
HAT-P-13c causes 8.5 s 
light-time effect (LITE) and perturbations in the orbit of HAT-P-13b.
On short ($\approx$1 yr) time scales, the LITE is dominant.
But the expected LITE is smaller than the ambiguity of our transit times
by a factor of 5, and therefore there is no chance for a positive detection
at this level of accuracy.

\begin{figure}
\includegraphics[width=8cm]{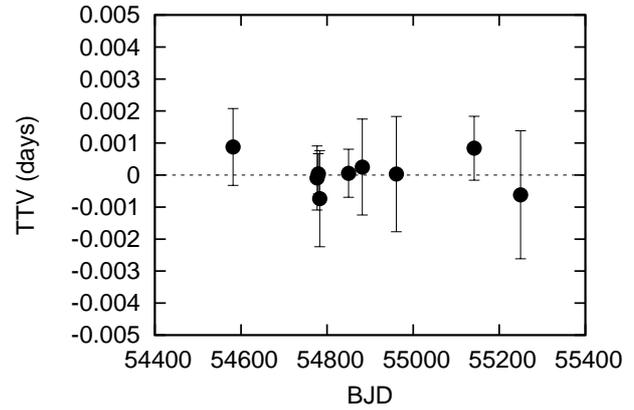}
\caption{Transit Timing Variation of HAT-P-13b.}
\end{figure}

\section{Summary}

\begin{itemize}
\item{} A multisite campaign has been organised to observe HAT-P-13 around the expected
transit of HAT-P-13c. Two transits of HAT-P-13b were also observed.

\item{}
HAT-P-13c was not observed to transit. We concluded that HAT-P-13c is not a
transiting planet with 75\%{} significance.

\item{} The refined period of HAT-P-13b is 2.916293$\pm$0.000010 days.
The determined TTV is consistent with zero variation. 
\end{itemize}

\begin{acknowledgements}
This project has been supported by 
the Hungarian OTKA Grants K76816 and MB08C 81013, and the ``Lend\"ulet'' Young
Researchers' Program of the Hungarian Academy of Sciences. GyMSz was supported
by the `Bolyai' Research Fellowship of the Hungarian Academy of Sciences.
The 91cm telescope of the Serra La Nave station is supported by 
INAF Osservatorio Astrofisico di Catania, Italy.
We acknowledge assistance of the queue observers, Karzaman Ahmad from LNO and
Hsiang-Yao Hsiao from Lulin Observatory. ZsK acknowledges the support of the 
Hungarian OTKA grants K68626 and K81421.
\end{acknowledgements}

\end{document}